\documentclass[10pt, a4paper]{article}

\usepackage{styles/arxiv}         
\usepackage[utf8]{inputenc}

\usepackage[english]{babel}
\usepackage[T1]{fontenc}    
\usepackage{hyperref}       
\usepackage{url}            
\usepackage{booktabs}       
\usepackage{amsfonts}       
\usepackage{microtype}      
\usepackage{amsmath}
\usepackage{amssymb}        
\usepackage{amsthm}         
\usepackage{dsfont}         
\usepackage{float}
\usepackage{graphicx}
\usepackage{stmaryrd}       
\usepackage{tikz}
\usepackage{csquotes}

\usepackage{pgfplots}
\usepgfplotslibrary{colormaps}

\usepackage{todonotes}
\usepackage{caption}
\usepackage{subcaption}     

\usepackage{algorithm}
\usepackage{algpseudocode}

\usepackage{lipsum}

\usepackage{natbib}
\pgfplotsset{compat = newest}
\pgfplotsset{
    colormap={uni}{rgb255(0cm)= (0, 170, 220); rgb255(1cm)= (255, 20, 11)}
}

\definecolor{myblue}{RGB}{0, 170, 220}

\newcommand\cmp[3]{((#1^x)/((#3) * (x!)^#2))}

\title{Ranking Handball Teams from Statistical Strength Estimation}

\date{} 					

\author{
 Florian Felice\textsuperscript{*}\\
 Department of Mathematics\\
 University of Luxembourg\\
 \href{mailto:florian.felice@uni.lu}{\texttt{florian.felice@uni.lu}}\\
}

\begin{document}

\maketitle

\let\thefootnote\relax\footnotetext{\textsuperscript{*} Work not related to Amazon}

\begin{abstract}
In this work, we present a methodology to estimate the strength of handball teams using a statistical method.
We propose the use of the Conway-Maxwell-Poisson distribution to model the number of goals scored by a team as a flexible discrete distribution which can handle situations of non equi-dispersion.
From its parameters, we derive a mathematical formula to determine the strength of a team.
We propose a ranking based on the estimated strengths to compare teams across different championships.
Applied to female handball club data from European competitions over the season 2022/2023, we show that our new ranking can have an echo in real sports events and the results.
\end{abstract}

\section{Background and related work}\label{sec:intro}

Handball is a popular sport with growing interest across the world.
To date, there does not exist any official ranking tool to compare clubs and players performances, the only quantitative metrics available provided by the European and International Handball Federations (EHF and IHF) are coefficient ranks that compare countries (based on championships).
Handball also suffers from a lack of literature~\citep{saavedra_handball_2018}, in particular in the predictive or analytical fields.
In this work, we aim to establish a methodology to estimate the strength of a team via a statistical procedure.

Estimating the strength of a team has long been discussed in the literature, and particularly for football.
Rating methods often assume some probability distributions to represent the number of goals scored by a team.
Some methods are based on the Thurstone-Mosteller model \citep{thurstone_psychophysical_1927} or the Bradley–Terry model \citep{bradley_rank_1952} to model the outcome of a match, based on some probability distribution whose location parameter corresponds to the strengths of the modeled teams.
These popular techniques, however, assume that the underlying probability distribution is continuous which is, by nature, in contradiction with the structure of the majority of sports data.

The choice of the underlying probability distribution to represent the number of goals scored by a team also leads to debates.
\cite{reep_skill_1971} demonstrated that the Negative Binomial is a suitable distribution to model scores in several ball games.
\cite{maher_modelling_1982} however, argued that tests for goodness-of-fit plead in favor of the independent Poisson distribution to model football scores.
\cite{ley_ranking_2019} further investigated the idea of Poisson distributions and, based on a broad comparison of models, suggested the bivariate Poisson model \citep{karlis_analysis_2003} to represent the outcome of football games.
From the estimated parameter $\lambda$ obtained via Maximum Likelihood Estimation, they assume a structure from the parameter for team $i$ with opponent team $j$ as

\begin{equation}
  \log(\lambda_i) = \beta_0 + (r_i - r_j) + h \cdot \mathds{1}(\text{team } i \text{ playing at home})
\end{equation}

where $\beta_0 \in \mathbb{R}$ is a common intercept and $h > 0$ is the effect of playing at home.
The parameters $r_i > 0$ and $r_j > 0$ represent the abilities of team $i$ and $j$ that are used as estimation of team's strength.

In the context of handball, \cite{groll_prediction_2020} analyzed historical international games to determine the best probability distribution to model the number of goals scored in handball matches.
Given the level of under-dispersion observed, they concluded that the standard Poisson distribution cannot be used and a Gaussian distribution with low variance is the most appropriate.

In this article, we propose a method to derive a ranking based on handball teams strengths.
These strengths are obtained using the estimated parameters of an appropriate discrete probability distribution by means of maximum likelihood.
We define formulae to transform such statistical estimates into sports abilities and shall observe how mathematical expressions can translate into actual sports facts.
To illustrate our results, we apply our method to historical European female matches over the season 2022/2023 and obtain a ranking which is linked to the end of season standings.

Our work is organized as follows. In Section~\ref{sec:methodo}, we will compare methods from the existing literature with the Conway-Maxwell-Poisson distribution.
After motivating the use of this flexible discrete probability distribution, we will generate a metric representing the strength of a team.
In Section~\ref{sec:applications}, we will illustrate the results of the proposed methodology on female club data and propose a ranking of the best performing team based on statistical facts.
Finally, we will discuss next steps and future considerations in Section~\ref{sec:discussion} and conclude in Section~\ref{sec:conclusion}.

\section{Methodology}\label{sec:methodo}

In this section, we present the methodology for modeling handball data to represent the strength of a team.
We first justify why the classical Poisson distribution cannot be used as the underlying probability distribution.
As an alternative, we propose the Conway-Maxwell-Poisson distribution as a flexible probability distribution to estimate, from its parameters, the strength of a team.

\subsection{Non equi-dispersion from handball data}

When analyzing historical data from female handball matches, one can observe situations with non equi-dispersion.
We define the dispersion index $DI$ as the ratio between the expectation $\mathbb{E}(X)$ and the variance $\mathbb{V}(X)$ of a random variable:
\begin{equation}
    DI = \dfrac{\mathbb{E}(X)}{\mathbb{V}(X)}.
\end{equation}

When $DI < 1$, we are in the situation of over-dispersion, the variance being larger than the expectation.
When $DI > 1$, the variance is lower than the average which corresponds to under-dispersion.
The final situation where $DI = 1$ leads to equi-dispersion.

To measure such index for female handball data, we analyzed games over the season 2022/2023 in European championships and observed that the empirical mean $\hat{\mathbb{E}}(X) = 27.9$ is lower than the empirical variance $\hat{\mathbb{V}}(X) = 31.5$.
This leads to a dispersion index $DI = 0.88$, suggesting over-dispersion.
Therefore, aligned with conclusions from \cite{groll_prediction_2020}, the equi-dispersed Poisson distribution cannot be used to model scored goals during handball matches.

\subsection{Modelling handball games with Conway-Maxwell-Poisson}\label{sec:cmp_model}


As an alternative to the standard Poisson distribution, we consider the Conway-Maxwell-Poisson (CMP) distribution \citep{sellers_conway-maxwell-poisson_2022}.
It is a generalization of the common Poisson distribution, but can handle situations with under- and over-dispersion.
Its probability mass function is defined by

\begin{equation}
    \mathbb{P}(X = x | \lambda, \nu) = \dfrac{\lambda^{x}}{(x!)^{\nu}} \dfrac{1}{\sum_{j=0}^{\infty} \dfrac{\lambda^{j}}{(j!)^{\nu}}}.
\end{equation}

The parameter $\nu \geq 0$ represents the level of dispersion.
When $\nu = 1$, one retrieves the equi-dispersed Poisson distribution.
When $\nu < 1$, we are in the situation of over-dispersion while $\nu > 1$ represents under-dispersion.
Though it does not have an explicit interpretation, $\lambda > 0$ can be seen as a location parameter whose value gets closer to the mean as $\nu \rightarrow 1$.
Other special cases of the Conway-Maxwell-Poisson distribution approach the Bernoulli with parameter $\lambda /(1+\lambda )$ as $\nu \rightarrow \infty$ and the geometric distribution with probability of success $1-\lambda$ when $\lambda < 1$ and $\nu = 0$.
The distribution can thus be a good alternative to the classical Poisson distribution given its flexibility to handle different levels of dispersion.

To evaluate the goodness of fit of the distribution on handball data, we compare the CMP with the Gaussian and Negative Binomial distributions as mentioned in \cite{groll_prediction_2020}.
In Table~\ref{tab:loglike}, we report the estimated log-likelihood ($\hat{L}$) and the associated Akaike Information Criterion (AIC) estimated for the club of Metz Handball over the season 2022/2023.

\begin{table}[h]
  \centering
  \caption{Comparison of log-likelihood and AIC evaluated on scored goals by Metz Handball over season 2022/2023.
  }\label{tab:loglike}
  \begin{tabular}{ccc}
    \toprule
    Distribution & Log-likelihood & AIC \\
    \midrule
    Conway-Maxwell-Poisson & -127.36 & 258.72\\
    Gaussian & -127.39 & 258.78\\
    Negative Binomial & -127.66 & 259.32\\
    \bottomrule
  \end{tabular}
\end{table}

We observe from Table~\ref{tab:loglike} that, although the three distributions seem to similarly fit the data, the Conway-Maxwell-Poisson distribution minimizes the AIC.
Although the AIC aims to penalize complex distributions with numerous parameters to estimate, given that $k = 2$ for all three distributions, minimizing the AIC or maximizing the log-likelihood leads to the same conclusion.
One can also argue that, given the flexibility of the distribution, which can handle under-, equi- and over-dispersion situations, this Conway-Maxwell-Poisson distribution can be the most appropriate choice.
We also noted from our experiments that these results and conclusions also apply for other teams.

\begin{figure}[h]
  \centering
  \begin{tikzpicture}
    \begin{axis}[xlabel=$\lambda$,
        ylabel=$\nu$,
        zlabel=Empirical mean,
        x dir=reverse,
        x tick label style={draw=none},
        y tick label style={draw=none},
        z tick label style={draw=none}]
    \addplot3 [surf, mesh/rows=10, only marks, scatter]
    table[x=lambda, y=nu, z=emp_mean, col sep=comma] {figures/data/CMP_simu.csv};
    \end{axis}
\end{tikzpicture}
  \caption{Relation between CMP parameters $\lambda$ and $\nu$ and the empirical expectation}
  \label{fig:simu_cmp}
\end{figure}
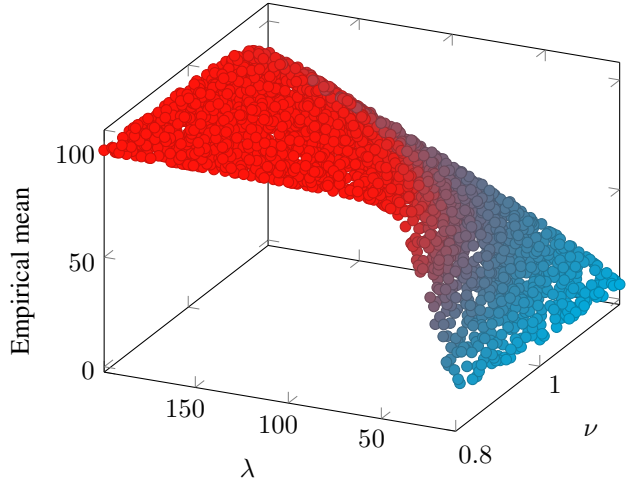

We represent in Figure~\ref{fig:simu_cmp} the relation between the empirical mean from a CMP distribution and its associated parameters $\lambda$ and $\nu$. We  notice a logarithmic relation between the parameter $\lambda$ and the empirical mean.
This relation is of particular interest in the next Section~\ref{sec:team_strength} when defining the team's strength.

\subsection{Estimation of team strengths}\label{sec:team_strength}

The strength of a team can be expressed by its ability to perform in attack and in defense.
We thus introduce different formulae to represent defense and attack strengths of a team.
We then define the overall strength of a team as a combination of attack and defense abilities.

\subsubsection{Defense strength}\label{sec:team_strength_defense}

Adopting the selected Conway-Maxwell-Poisson (CMP) distribution, we use its parameters to represent the strength of a team in defense.
The distribution of goals conceded by a team, denoted by $Y_d$, is assumed to follow a $CMP(\lambda_d, \nu_d)$, where the parameter $\lambda_d > 0$ can act as a location parameter and $\nu_d \geq 0$ as the dispersion parameter.
We then define the defense strength as

\begin{equation}\label{eq:defense_strength}
    s_d = \dfrac{\nu_{d}}{\log(\lambda_d)}.
\end{equation}

The strength of a team's defense is inversely proportional to the goals it concedes.
This is reflected in equation (\ref{eq:defense_strength}) in the sense that the higher the average number of conceded goals (i.e. the higher $\lambda_d$) the lower the strength $s_d$.
We notice the logarithmic transformation $\log(\lambda_d)$ to account for the relation with the empirical mean as mentioned and illustrated in Figure~\ref{fig:simu_cmp}.
On the other hand, we want to penalize for irregularities of a team, therefore we want the parameter $\nu_d$ to be as large as possible corresponding to under-dispersion.
We can thus interpret formula (\ref{eq:defense_strength}) as follows: a team is a strong defender if it constantly concedes few goals during matches.

\subsubsection{Attack strength}\label{sec:team_strength_attack}

We also assume that the distribution of scored goals follows a CMP distribution, $Y_a \sim CMP(\lambda_a, \nu_a)$.
A team is considered strong in attack if the average number of scored goals is large.
The logic can therefore be considered as the inverse from equation (\ref{eq:defense_strength}).
We define the attack strength of a team as

\begin{equation}\label{eq:attack_strength}
    s_a = \dfrac{\log(\lambda_a)}{\nu_a}
\end{equation}
where  the location parameter $\lambda_a$ is  used as the numerator to show that a high number of goals scored on average increases the attach strength.
The parameter $\nu_a$ is used as a penalty as we expect teams to have regular performances over the season, but they should also be capable of occasionally scoring numerous goals when facing weaker teams.
From this mixed requirements, the dispersion parameter is used as the denominator.

\subsubsection{Global strength}\label{sec:team_strength_global}

A team is considered strong when it can perform well in attack and defense.
We consider the overall strength of a team as the combination of attack and defense strengths by

\begin{equation}\label{eq:overall_strength}
  s = s_a \cdot s_d = \dfrac{\log(\lambda_a) \cdot \nu_{d}}{\nu_{a} \cdot \log(\lambda_d)}.
\end{equation}

We observe that a high score for overall strength can be driven by two factors.
On the one hand, the team should have a high average of scored goals while demonstrating constant defense performances over time.
On the other hand, a team should be able to adapt its attack strategies to teams and be able to score more than expected, taking their opponent by surprise.
It should also be able to prevent conceding too many goals and have a low average of conceded goals.

In other words, the goal difference in the competitions ranking should be as large as possible.
This can usually be verified in different competitions where leading teams tend to have a high difference (+229 goals for Metz Handball in the French female championship at end of season 2022/2023 or +257 for Vipers Kristiansand in Norway) while teams at the bottom of the season standings have highly negative goal differences (-107 for Toulon Métropole Var Handball in France for the same season or -170 for Volda in Norway.)

We can now note the importance of the nonlinear transformation for $\lambda_a$ and $\lambda_d$.
Given the logarithmic rate of these parameters, one may have to record a much higher average of scored goals to distinguish itself from other teams.
Indeed, considering the slope of the strength with respect to scored goals as $\frac{\partial s}{\partial \lambda_a} \propto \frac{1}{\lambda_a}$, as teams get stronger, $\lambda_a$ gets higher and differentiators between teams become marginal since $\lim_{\lambda_a \rightarrow \infty} (\frac{1}{\lambda_a}) = 0$. On the contrary, any improvement in defense performances can lead to more important improvements in the overall strength.
Because $\frac{\partial s}{\partial \lambda_d} \propto \lambda_d$, the strength will grow linearly as the average of conceded goals decreases.

These statements can also have an echo in sports terms.
It is common knowledge for handball players and coaches that the best way to improve a team's performance is to start by improving defense.

\section{Illustrative applications}\label{sec:applications}

\begin{figure}[h]
  \centering
  \begin{tikzpicture}
    \begin{axis}[
    height=7cm,
    width=10cm,
    xmin=10,
    xmax=50,
    xlabel = Number of goals,
    ylabel = Frequency,
    legend style={nodes={scale=0.7, transform shape}},
    x tick label style={draw=none},
    y tick label style={font=\footnotesize, draw=none},
    ]

   \addplot[
       black,
       fill=lightgray,
       hist=density,
       hist/bins=20,
   ] table[y=y] {figures/data/dist.csv};

   \addplot[domain={10:50}, samples=40, dashed] {\cmp{250.45803272774776}{1.6390174920550273}{4193921903904605339648/3}};
   \addlegendentry{Empirical}
   \addlegendentry{$CMP(286.46, 1.64)$}

   \end{axis}
\end{tikzpicture}
  \caption{Histogram of goals scored by Metz Handball over season 2022/2023 vs. theoretical CMP distribution}
  \label{fig:emp_theo}
\end{figure}
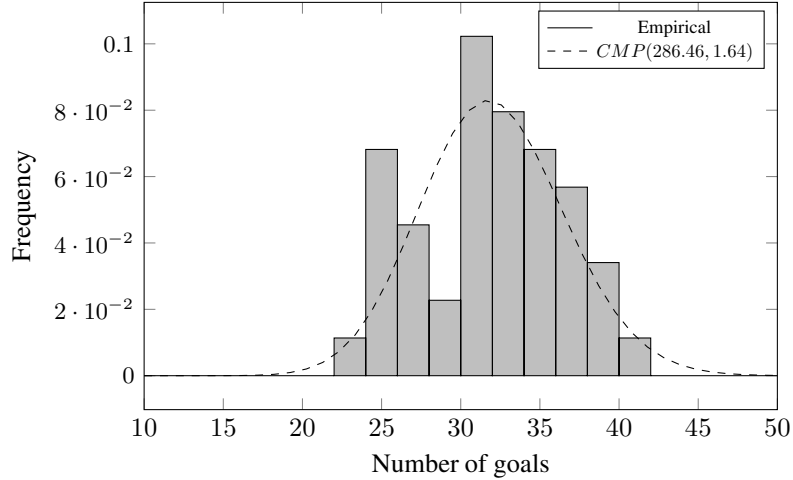

As illustrated in Table~\ref{tab:loglike} from Section~\ref{sec:cmp_model}, the CMP distribution seems to be the most appropriate choice to model goals scored during a handball match.
We plot in Figure~\ref{fig:emp_theo} the histogram of scored goals over the season 2022/2023 for the female club of Metz Handball (France) and compare with the fitted theoretical CMP distribution.
Furthermore, we estimate the strength parameters for European female clubs and display the ranking in Table~\ref{tab:top_strength}.
The estimations are derived from all games over the season 2022/2023, played in all first division female competitions (from friendly games, to the regular championships and champions' league).

\begin{table}[h]
  \centering
  \caption{Top 10 strongest female teams in Europe for season 2022/2023.
  }\label{tab:top_strength}
  \begin{tabular}{cccccc}
    \toprule
    Team & Avg. scored & Avg. conceded & Attack strength & Defense strength & Strength \\
    \midrule
    Győri Audi ETO KC & 33.32 & 24.32 & 3.49 & 3.16 & 11.00\\
    Vipers Kristiansand & 37.62 & 26.38 & 3.57 & 3.07 & 10.96\\
    Podravka Vegeta & 30.50 & 21.75 & 3.39 & 3.21 & 10.89\\
    Metz handball & 33.58 & 24.00 & 3.47 & 3.12 & 10.85\\
    Team Esbjerg  & 33.33 & 24.67 & 3.48 & 3.11 & 10.83\\
    SG BBM Bietigheim & 34.63 & 25.21 & 3.54 & 3.05 & 10.80\\
    HC Dunajsk\`a Streda & 29.37 & 22.62 & 3.38 & 3.15 & 10.63\\
    Herning-Ikast Håndbold & 28.71 & 23.29 & 3.39 & 3.13 & 10.61\\
    DVSC Schaeffler & 30.83 & 24.11 & 3.39 & 3.12 & 10.59\\
    CSM București & 33.13 & 25.83 & 3.48 & 3.05 & 10.58\\
    \bottomrule
  \end{tabular}
\end{table}

We can observe that the teams considered the strongest are mostly strong competitors in the female European Champions League.
In particular, the top clubs Győri Audi ETO KC and Vipers Kristiansand were part of the EHF final four in June 2023 and the latter club won the competition.
Other clubs are leading their championships in their respective countries.

We also notice that in Table~\ref{tab:top_strength}, even though the ranking is sorted by the overall estimated strength, the average number of goals scored and conceded seems to follow a hierarchy.
The top clubs clearly show a high average number of scored goals and a relatively lower number of conceded goals.
Some clubs (e.g. SC BBM Bietigheim) can record higher scored goals and still be ranked lower (e.g. compared to Team Esbjerg) due to lower defense ability but also more irregularities.
Indeed, such teams suffer from a higher value for $\nu_a$ or lower value for $\nu_d$ suggesting irregularities in attack or defense and penalizing them in the final strength ranking.
This justifies the need of formulae (\ref{eq:defense_strength}) and (\ref{eq:attack_strength}) instead of purely relying on average scored goals.

\section{Discussion}\label{sec:discussion}

Our proposal offers an estimation of attack and defense strengths in order to rank teams and generate features that can be informative and meaningful in subsequent modelling tasks.
Provided that one has access to such data, the presented exercise can be extended to other objectives such as estimation of player abilities or generalized to other sports.

\subsection{From team strengths to player abilities}\label{sec:player_ability}

Using more granular data (not publicly available) on player performances for each game and over several seasons, one can also estimate the attack strength of a player.
Considering that the data will most likely also suffer from under- or over-dispersion, the CMP distribution seems a good choice to fit the number of scored goals by a player.
Using formula (\ref{eq:attack_strength}), we can therefore estimate the attack strength of an individual player.
Not focusing only on goals scored, playing ability could also include components such as passing ability and combine scoring and pass abilities as a global attack strength. Accessing data such as interception, successful blocks (e.g. faults with no penalty such as yellow card, 2 minutes penalty, etc.), the defense ability can be modeled in a similar fashion and one can derive a defense ability at player level.

Therefore, combining attack and defense abilities as defined by equation (\ref{eq:overall_strength}), one can estimate the individual abilities and derive a ranking.
Such ranking can help subsequent modelling exercises by adding informative variables regarding the strength of the individual players and not only the global strength of a team.
Additionally, the individual ranking can be used as a new source of information for team managers to assess the potential abilities of a player when recruiting.
Indeed, one can obtain a time dependent ranking and observe the evolution of a player over several seasons.
This can further lead to forecasting exercises in order to identify players with high potential to be added to the squad.

\subsection{Generalization to other sports from Conway-Maxwell-Poisson distribution}\label{sec:general_ability_cmp}

Modelling sports requires to rely on discrete distributions though the issue of over- or under-dispersion is a recurrent problem \citep{karlis_bayesian_2008, van_bommel_home_2021}.
Given the similar constraints as in the present work, one can replicate the discussed logic on other sports' data.
The methodology from \cite{ley_ranking_2019} can be merged with our proposed methodology in order to obtain football team abilities based on a distribution that can handle the problem of under-dispersion.
One can thus define new rankings and generate new informative features to include in predictive Machine Learning models.
Using the framework of Statistically Enhanced Learning (SEL) \citep{felice_statistically_2023} one can include such generated features in the feature set to improve the predictive model.

\section{Conclusion}\label{sec:conclusion}

Handball is a fast-paced sport of which goals cannot be analyzed via standard count distributions due to the problem of under- or over-dispersion.
We showed that, using an appropriate probability distribution, one can define meaningful statistical estimates that approximate the strength of a team.

The proposed methodology allows to generate very informative features that can be included in predictive models in the spirit of Statistically Enhanced Learning.
They can also offer the possibility to consider data-driven analyses of a team's performance to later support team managers in their definition of sports strategies.
With access to more granular data, this methodology can be adapted to the estimation of player abilities and offer tools to allow coaches take data-driven decisions in their recruitment processes.



\bibliography{references2}

\end{document}